\documentclass[12pt,a4paper]{article}

\usepackage[T1]{fontenc}
\usepackage[utf8]{inputenc}
\usepackage{authblk}
\usepackage{amsfonts}
\usepackage{mathtools}
\usepackage{slashed}
\usepackage{amsmath,amssymb}
\usepackage{graphicx,xcolor}
\usepackage{cite}
\usepackage{hyperref}

\definecolor{refcol}{rgb}{0.9,0.1,0.1}
\hypersetup{colorlinks=true,linkcolor=blue,citecolor=refcol,urlcolor=cyan,linktocpage}
  
\usepackage[capitalize]{cleveref}
\usepackage{tikz}
\usetikzlibrary{shapes}
\usetikzlibrary{plotmarks}

\usepackage{setspace}

\usepackage[left=2.5cm,right=2.5cm,top=2.5cm,bottom=2.5cm]{geometry}
\allowdisplaybreaks
\newcommand{\be}{\begin{equation}}
\newcommand{\ee}{\end{equation}}
\newcommand{\bea}{\begin{eqnarray}}
\newcommand{\eea}{\end{eqnarray}}

\def\XXint#1#2#3{{\setbox0=\hbox{$#1{#2#3}{\int}$ }
\vcenter{\hbox{$#2#3$ }}\kern-.6\wd0}}

\newcommand{\Qp}{\mathbb{Q}_p}
\newcommand{\Zp}{\mathbb{Z}_p}

\begin{document}

\begin{titlepage}
\thispagestyle{empty}

\title{
{\huge\bf Phase operator on $L^2(\Qp)$ and}\\ 
{\huge\bf the zeroes of Fisher and Riemann}
}

\author{{\bf Parikshit Dutta}${}^{1}$\thanks{{\tt parikshitdutta@yahoo.co.in}}\hspace{8pt} {\bf and} \hspace{2pt}  
                    {\bf Debashis Ghoshal}${}^2$\thanks{{\tt dghoshal@mail.jnu.ac.in}} \\  
\hfill\\
${}^1${\it Asutosh College, 92 Shyama Prasad Mukherjee Road,}\\
{\it Kolkata 700026, India}
\hfill\\              
${}^2${\it School of Physical Sciences, Jawaharlal Nehru University,}\\
{\it New Delhi 110067, India}
}

\date{%
%
\begin{quote}
\centerline{{\bf Abstract}}
{\small
The distribution of the non-trivial zeroes of the Riemann zeta function, according to the Riemann hypothesis, is tantalisingly similar to 
the zeroes of the partition functions (Fisher and Yang-Lee zeroes) of statistical mechanical models studied by physicists. The resolvent 
function of an operator akin to the phase operator, conjugate to the number operator in quantum mechanics, turns out to be important in this 
approach. The generalised Vladimirov derivative acting on the space $L^2(\Qp)$ of complex valued locally constant functions on the 
$p$-adic field is rather similar to the number operator. We show that a `phase operator' conjugate to it can be constructed 
on a subspace $L^2(p^{-1}\Zp)$ of $L^2(\mathbb{Q}_p)$. We discuss (at physicists' level of rigour) how to combine this for all primes 
to possibly relate to the zeroes of the Riemann zeta function. Finally, we extend these results to the family of Dirichlet $L$-functions, 
using our recent construction of Vladimirov derivative like pseudodifferential operators associated with the Dirichlet characters.
}
\end{quote}
}

\vfill


\end{titlepage}

\thispagestyle{empty}
\maketitle


\newpage

\section{Introduction}\label{sec:Introd}
The statistical distribution of the zeroes of the Riemann zeta function, and the related family of Dirichlet $L$-functions, qualitatively 
resemble the eigenvalue distribution of a random ensemble of unitary matrices\cite{montgomery, anecdote,Odlyzko}. It is also
reminiscent of the distribution of zeroes of partition functions of statistical models. The latter observation is the motivation to search 
for a suitable model in physicists' approach to the problem---the literature is vast, however, see e.g.,
Refs.\cite{Spector1990,Julia1990,Julia1994,Bakas1991,KnaufNumSpin,KnaufSpinRZ}, the review \cite{Schumayer:2011yp} and 
references therein. This resemblance may be an important guide as the zeroes of the partition function of many systems, by the 
Yang-Lee type theorems \cite{Itzykson1991SFT}, all lie parallel to the imaginary axis (or on the unit circle). These zeroes are called 
Yang-Lee zeroes or Fisher zeroes, depending upon whether the partition function is viewed as a function of the applied external field, 
e.g., magnetic field, or of $\beta=1/(k_BT)$, the inverse temperature. The arithmetic or primon gas of Refs.
\cite{Julia1990,Bakas1991,Julia1994} and the number theoretic spin chain of Refs.\cite{KnaufNumSpin,KnaufSpinRZ}, in particular, 
proposed interesting models for which the partition functions are directly related to the Riemann zeta function. In this approach the 
non-trivial zeroes of the zeta function are to be identified with the Yang-Lee or Fisher zeroes. 

Motivated by these, we shall propose a statistical model and compute its partition function. The idea is again to associate the spectrum 
of an operator with the Fisher zeroes of the partition function. In addition, however, we shall study the spectrum of some relevant operators
of these models. The systems that relate to the $L$-functions of our interest can be thought of as spins in an external magnetic field. 
Since the spectrum of a Hamiltonian of this type of spin systems is discrete (the spins being integer/half-integer valued) this operator 
is similar to the number operator of an oscillator. A phase operator that is conjugate to this will be a new ingredient in our investigation. 
The construction of a phase operator which is truly canonically conjugate to the number operator is a subject of long-standing quest that 
may not be completely closed yet. Nevertheless, several different ways to define the phase operator have been proposed, for example, 
Refs.\cite{SusskindGlo,PhaseOpRevCN,GarrisonWong,GalindoPhase,PeggBarnett,BuschEtAl,MaRhodes} is a partial list. In particular, 
we shall investigate two ways of defining it for the spin models corresponding to the family of $L$-functions. In the first construction, we  
follow Ref.\cite{BuschEtAl}, where the authors propose an operator by directly constructing eigenstates of phase for a system with a 
discrete spectrum. The second approach is motivated by the proposal in \cite{GalindoPhase}. We shall argue that there are enough 
hints in these proposals to understand the correspondence between the spectrum of these operators and the zeroes of the partition 
function. 

In the following, we shall first review (in \cref{sec:Ising}) some of the relevant arguments and results from the cited references, in the 
context of a simple spin system on a one-dimensional lattice. In \cref{sec:RiemannZ}, after recalling some properties of the Riemann 
zeta function and our earlier work on its relation to operators on the Hilbert space of complex valued functions on the $p$-adic number 
field $\Qp$ \cite{Chattopadhyay:2018bzs,Dutta:2020qed}, we elaborate on a proposal to view it as a statistical model of spins. In 
\cref{ssec:RZPhOp,ssec:AnotherPhOp} we detail two constructions of the phase operators for the spin model for the Riemann zeta 
function, which are then extended to the family of Dirichlet $L$-functions in \cref{sec:DirichletL}.  
			 		
\section{Quantum spins in external field}\label{sec:Ising}
Spin models in one dimension are among the simplest statistical models, yet they offer an arena rich enough to experiment, before 
considering more complicated systems. The variables are `spins' $s_n$ at lattice points $n\in\mathbb{Z}$ or $n\in\mathbb{N}$ that 
can take $(2j+1)$ values $\{-j,-j+1,\cdots,j-1,j\}$ in the spin-$j$ representation. In models of magnetism, these spins interact locally, 
usually with the nearest neighbours. In addition, one may turn on an external magnetic field. 

Let us digress to recall the properties of a simpler model, the Ising model, in which the {\em classical} spins take one of two possible 
values $\pm 1$ and the total Hamiltonian is $H = - J\sum_n s_n s_{n+1} - B\sum_n s_n$, where $J$ is the strength of interaction 
($J>0$ being ferromagnetic and anti-ferromagnetic otherwise) and the second term arises from an interaction with an external magnetic 
field $B$. The partition function (in the absence of an external field) of an Ising system of size $L$ at inverse temperature $\beta$ is 
\begin{equation*}
Z(\beta) \equiv \text{Tr } e^{-\beta H} = \sum_{\left\{s_n\right\}} \exp \left(\beta J\sum_{n=1}^{L-1} s_n s_{n+1}\right)
\end{equation*}
In this simple case, one may also change variables to $\sigma_n  \equiv \sigma_{\langle n-1,n\rangle} = s_{n-1} s_n$ 
associated to the edges $\langle n-1, n\rangle$ joining nearest neighbours. Evidently, $\sigma_n = \pm 1$ as well. Thus
\begin{eqnarray*}
Z(\beta) &=& 2\sum_{\left\{\sigma_n\right\}} \exp \left( \beta J \sum_{n=2}^{L} \sigma_n\right) \\
&=& 2 \sum_{\sigma_n} \left\langle \sigma_{2},\sigma_{3},\cdots \right| \exp\left( \beta J\sum_{j}S_{j}
\right) \left| \sigma_{2},\sigma_{3},\cdots \right\rangle
\end{eqnarray*}
where we have defined vectors $|\sigma_n\rangle$ in a (two-dimensional) Hilbert space corresponding to the edge $\langle n-1, n\rangle$ 
and $S_n$s are spin operators such that $S_n \left| \sigma_n \right\rangle = \sigma_n \left| \sigma_n \right\rangle$. A generalisation of this 
model allows the coupling constants $J$ to be position dependent, so that the Hamiltonian is $H = - \sum_{n} J_n \sigma_n$ and
\begin{equation*}
Z(\beta) = 2 \sum_{\sigma_n} \big\langle \sigma_{2},\sigma_{3},\cdots \big| e^{\beta \sum_n J_n S_n} 
\big| \sigma_{2},\sigma_{3},\cdots \big\rangle
\end{equation*}
is the canonical partition function of the generalised model at the temperature $k_BT=1/\beta$. 

We would like to consider the general case where the spins to be valued in the spin-$j$ representation of $\mathfrak{su}(2)$. Although we
seek a partition function of the form as above, the general spin case is cannot be realised as an Ising type model, rather it will be a model 
of spins in an external local magnetic field $B_n$ at site $n$. It will be useful to think of $S_n$ to be the third component $S_{n3}$ of the 
$\mathfrak{su}(2)$ spin operators on the edge/site, the others being $S_{n\pm}$. The vectors $| \sigma_{2},\sigma_{3},\cdots\rangle = 
|\sigma_{2}\rangle \otimes |\sigma_{3}\rangle \otimes\cdots$ belong to the product space.The interaction $\sim\mathbf{B}\cdot\mathbf{S}$
between the spin (magnetic moment to be precise) and the external field (assumed to be along the $z$-direction) described by the 
Hamiltonian $H = - \sum_{n} B_n \sigma_n$ leads to the partition function
\begin{equation*}
Z(\beta) = \sum_{\sigma_n} \big\langle \sigma_{2},\sigma_{3},\cdots \big| e^{\beta \sum_n B_n S_n} 
\big| \sigma_{2},\sigma_{3},\cdots \big\rangle
\end{equation*}
at the temperature $k_BT=1/\beta$. Our objective is to obtain an identity for the partition function for this model. To this end, we shall seek 
an operator that, in a certain well defined sense, is formally canonically conjugate to the $z$-component $S_{n3}$ of the spin operator at 
site $n$. There are well known difficulties in defining such an operator, however, we shall see that one needs to make a much weaker 
demand. 

In this context, it is useful to remeber the Schwinger oscillator realisation of the algebra $\mathfrak{su}(2)$ in terms of a pair of bosonic 
creation/anhilation operators $(a_1^\dagger, a_1, a_2^\dagger, a_2)$ at each edge, where we drop the edge index for the time being. 
Then $S_{+}=a_{1}^{\dagger}a_{2}$, $S_{-}=a_{2}^{\dagger}a_{1}$ and the third component is the difference of the number operators 
$S_{3} = \frac{1}{2}(n_1 - n_2) = \frac{1}{2} \left(a_{1}^{\dagger}a_{1} - a_{2}^{\dagger} a_{2}\right)$. One can formally introduce the 
\emph{phase operator} $\Phi=\frac{1}{2} (\phi_{1} - \phi_{2})$ such that $[\phi_a,n_b] = i \delta_{ab}$, however, there are several 
mathematical difficulties in defining the above\cite{SusskindGlo,PhaseOpRevCN}. We will now review an explicit construction to show 
how one can still work around this problem. 

\subsection{Phase operator via phase eigenstates}\label{ssec:IsingPhaseEigen}
Let us label the eigenstates in the spin-$j$ representation of $S_{3}$ as $|m\rangle$, for $m=-j,\cdots,j$. One can define an eigenstate 
of phase as a unitary transform of these states as
\begin{align}
|\phi_{k}\rangle &= \frac{1}{\sqrt{2j+1}} \sum_{m=-j}^{j} e^{-im\phi_{k}B} |m\rangle \nonumber\\
\text{where, }\; \phi_{k} &= \frac{2\pi k}{B(2j+1)}, \qquad k=-j,\cdots,j \label{IsingPhState}
\end{align}
are the eigenvalues of the phase. The phase eigenstates satisfy
\begin{equation}
\langle\phi_{k'}|\phi_{k}\rangle =
\frac{1}{2j+1}\sum_{m=-j}^{j}e^{-imB(\phi_{k}-\phi_{k'})} = \delta_{k,k'} \label{IsingPhOrtho}
\end{equation}
and thus provide an orthonormal basis of the Hilbert space of states.

In terms of these, we may define the the `phase operator' through spectral decomposition as
\begin{equation}
\hat{\phi} = \sum_{k=-j}^j \phi_{k}\, |\phi_{k}\rangle \langle\phi_{k}|  \label{IsingPhOp}
\end{equation}
We shall now show that it transforms covariantly when conjugated by $e^{\beta B S_{3}}$. This works for special values of $\beta$ since, an 
eigenvalues $\phi_{k}$ of $\hat{\phi}$, being angle-valued, is only defined modulo $2\pi/B$. In order to see this, we note that
\begin{equation*}
e^{-\beta B S_{3}} \hat{\phi} \,e^{\beta B S_{3}} 
= \sum_{k=-j}^{j} \frac{\phi_{k}}{2j+1} \sum_{m=-j}^{j} \sum_{m'=-j}^{j} e^{-im(\phi_{k}-i\beta)B + im'(\phi_{k}-i\beta)B}
|m\rangle \langle m'| \nonumber
\end{equation*}
There are two cases to consider. The first is trivial: for $\beta=0$ or any of its periodic images $i\beta=\frac{2\pi n}{B}$ ($n\in\mathbb{Z}$) in
the complex $\beta$-plane, the RHS is the phase operator $\hat{\phi}$. More interestingly, if $i\beta$ takes any of the specific discrete values 
$\frac{2\pi k'}{B(2j+1)} + \frac{2\pi n}{B}$, where $k' = -j,\cdots,j$ (but $k'\ne 0$) and $n\in\mathbb{Z}$, i.e., $i\beta$ is a difference between the 
phase eigenvalues (mod ${2\pi}/{B}$), then $\phi_{k} - i\beta$ is again an allowed eigenvalue of the phase operator $\pmod {2\pi/B}$. In this 
case, we can add and subtract $i\beta$ to the eigenvalue $\phi_k$ and use the completeness of basis, to find
\begin{equation}
e^{-\beta BS_{3}}\hat{\phi}\, e^{\beta BS_{3}} = \hat{\phi} + i\beta\:
\text{ only for }\, 0\ne\beta = -\frac{2\pi i j}{B(2j+1)}, \cdots,\frac{2\pi i j}{B(2j+1)}\: \Big(\text{mod } \frac{2\pi}{B}\Big)  
\label{IsingPhCommutator}
\end{equation}
This is called a {\em shift covariance relation} \cite{Busch:2013}. It may also be rewritten as a commutator 
\begin{equation*}
\big[\hat{\phi}, e^{\beta BS_{3}}\big] = i\beta\, e^{\beta BS_{3}}\: \text{ only for }\, 
0\ne\beta = -\frac{2\pi i j}{B(2j+1)}, \cdots,\frac{2\pi i j}{B(2j+1)}\: \Big( \text{mod } \frac{2\pi}{B}\Big)  
\end{equation*}
i.e., at special values of the inverse temperature.

To summarise, we find that $\hat{\phi}$ in \cref{IsingPhOp} satisfies shift covariance, alternatively, though somewhat loosely, it is `canonically 
conjugate' to $S_{3}$ only for a special set of an infinite number of {\em imaginary} values of $\beta$, all on the line $\text{Re}\,\beta = 0$ as 
above. At $\beta=0$ (mod $\frac{2\pi}{B}$), the commutator is trivial\footnote{It is also reflected in the resolvent of the phase operator, as we 
shall see in the following.}. 

In passing, it is instructive to take the trace of the `canonical commutator'. The left hand side evidently vanishes, since the vector space of 
states is finite, namely $(2j+1)$, dimensional. On the right hand side, the trace of $e^{-\beta H}$, the partition function which vanishes, being 
a sum over the roots of unity. Thus the values of $\beta$ for which \cref{IsingPhCommutator} is valid must also satisfy the condition $\text{Tr }
e^{-\beta H}=0$. This means that mod ${2\pi}/{B}$, these values of  $i\beta\ne 0$ for which the partition function has a zero are same as that 
of the eigenvalues of $\hat{\phi}$.

The resolvent of the exponential of the phase operator at a single site (as a function of $z=e^{i\phi}$) is
\begin{equation*}
\hat{R}[\hat{\phi}](\phi) =  \left(1 - e^{-i\phi} e^{i\hat{\phi}}\right)^{-1}
\end{equation*}
and its trace is
\begin{equation*}
\mathrm{Tr } \left(\hat{R}[\hat{\phi}](\phi)\right)
= \sum_{k=-j}^j \Big\langle \phi_{k} \Big| \frac{1}{1 - e^{i(\hat{\phi} - \phi)}} \Big| \phi_{k}\Big\rangle\nonumber\\
= \sum_{k=-j}^{j}\frac{1}{1-e^{i(\phi_k - \phi)}}
\end{equation*}
On the other hand, the partition function at a single site $Z_1(\beta) = \text{Tr}\, e^{-\beta H} = \sum_m e^{\beta B m}$ vanishes at special 
values of the inverse temperature $\beta=\frac{2\pi mi}{B(2j+1)}$ (mod $\frac{2\pi}{B}$) where $m\in\{-j,\cdots,j\}$ but $m\ne 0$. These zeroes 
of the partition function in the complex $\beta$-plane are called {\em Fisher zeroes}. At precisely these values, the resolvent function develops 
poles. 

\section{The case of Riemann zeta function}\label{sec:RiemannZ}
Before we get to our main goal to interpret the Riemann zeta function as a partition function, let us briefly recall some of its relevant
properties. Originally defined by the analytical continuation of the series
\begin{equation}
\zeta(s) = \sum_{n=1}^\infty \frac{1}{n^s} = \prod_{p\,\in\,{\mathrm{primes}}} \frac{1}{\left(1 - p^{-s}\right)},
\qquad \mathrm{Re}(s) > 1 
\label{EulerSum}
\end{equation}
to the complex $s$-plane by Riemann, the zeta function has a set of equally spaced zeroes at negative even integers $-2n$, 
$n\in\mathbb{Z}$ called its trivial zeroes. More interestingly, it has another infinite set of zeroes, which, according to the 
\emph{Riemann hypothesis} lie on the \emph{critical line} ${\mathrm{Re}}(s) =\frac{1}{2}$. The related Riemann $\xi$-function 
(sometimes called the \emph{symmetric zeta-function}) and the {\em adelic zeta function} share only the latter (non-trivial) zeroes 
with \cref{EulerSum} (i.e., the set of trivial zeroes are absent in the following functions)
\begin{equation}
\xi(s) = \frac{1}{2}s(s-1)\zeta_{\mathbb{A}}(s) 
= \frac{1}{2}s(s-1)\pi^{-\frac{s}{2}}\Gamma\left(\frac{s}{2}\right)\zeta(s)  
\label{SymmZ}
\end{equation}
both of which satisfy the reflection identity $\xi(s)=\xi(1-s)$, respectively, $\zeta_{\mathbb{A}}(s)=
\zeta_{\mathbb{A}} (1-s)$, derived from a similar identity for the original zeta function. The former is a holomorphic function while 
the latter,  $\zeta_{\mathbb{A}}(s)$, is meromorphic.

The non-trivial zeroes of $\zeta(s)$ (which are the only zeroes of $\xi(s)$ and $\zeta_{\mathbb{A}}(s)$) conjecturally on the critical line, 
seem to occur randomly, although they are found to be correlated in the same way as the eigenvalues of a Gaussian ensemble of 
$N\times N$ hermitian or unitary matrices in the limit $N\to\infty$ \cite{montgomery,anecdote,Odlyzko}. Starting from Hilbert and P\'{o}lya,
it has long been thought that these zeroes correspond to the eigenvalues of an operator, that is self-adjoint in an appropriately defined sense. 
A direct analysis of the spectrum of the purported operator may lead to a proof of the Riemann hypothesis. Despite many ingenious efforts, 
an operator has not yet been found. In Ref.\cite{Chattopadhyay:2018bzs}, in a larger collaboration, we attempted to find a suitable operator 
by \emph{assuming the validity of the hypothesis}, specifically, by assuming that the zeroes \emph{are} the eigenvalues of a unitary matrix 
model (UMM). We found that the partition function can be expressed as the \emph{trace} of an operator on the Hilbert space of complex 
valued locally constant Bruhat-Schwarz functions supported on a compact subset $p^{-1}\Zp$ of the $p$-adic field $\Qp$. This was 
achieved in two steps. First a UMM was constructed for each prime $p$ corresponding to the Euler product form in \cref{EulerSum}. These 
(as well as a UMM for the trivial zeroes) were combined to define the random matrix model. In this  paper, we shall use some of the 
technology that were useful in \cite{Chattopadhyay:2018bzs}, however, our goal will be different. 

We begin by expanding the prime factors in the Euler product form of the zeta function 
\begin{equation}
\zeta(s) = \prod_{p\,\in\,{\mathrm{primes}}} \frac{1}{\left(1 - p^{-s}\right)}
= \prod_{p\,\in\,{\mathrm{primes}}} \sum_{n^{(p)}=0}^\infty p^{-s n_{(p)}}, \qquad {\mathrm{Re }}(s) > 1 
\label{EulerExpand}
\end{equation}
The factor
\begin{equation}
\zeta_p(s) = \frac{1}{\left(1 - p^{-s}\right)}\:\: {\text{for a fixed prime }} p \label{Zetap}
\end{equation}
is sometimes called the \emph{local} zeta function at $p$. It can be thought of as a complex valued function on the field 
$\Qp$ (of $p$-adic numbers). The prefactor 
$\zeta_{\mathbb{R}}(s) = \pi^{-{s\over 2}}\Gamma\left({s\over 2}\right)$ 
in \cref{SymmZ} is known as the \emph{local} zeta functions corresponding to $\mathbb{R}$ (of real numbers). It is the Mellin transform 
of the Gaussian function $e^{-\pi x^2}$. In an exactly analogous fashion, $\zeta_p(s)$ in \cref{Zetap} is the Mellin transform of the 
equivalent of the Gaussian function (in the sense of a function that is its own Fourier transform) on $\Qp$.

We can express the sum in \cref{EulerExpand} as the trace of an operator. To this end, let us recall that the space of (mean-zero) square 
integrable complex valued functions on $\Qp$ is spanned by the orthonormal set of Kozyrev wavelets $\psi^{(p)}_{nml}(\xi)\in\mathbb{C}$
(for $\xi\in\Qp$), which have compact  support in $\Qp$\cite{Kozyrev:2001}. In $p$ segments (of equal Haar measure) its values are the 
$p$-th roots of unity. They are analogous to  the generalised Haar wavelets, with the labels $n$, $m$ and $l$ referring to scaling, 
translation and phase rotation. Interestingly, the Kozyrev wavelets are eigenfunctions of an operator with eigenvalue $p^{\alpha(1-n)}$ 
\begin{equation}
D_{(p)}^\alpha \psi_{n,m,l}^{(p)} (\xi) = p^{\alpha(1-n)} \psi_{n,m,l}^{(p)} (\xi) \label{VDonKozy}
\end{equation}
where, the pseudodifferential operators $D_{(p)}^\alpha$, called the \emph{generalised Vladimirov derivatives}, are defined by the 
following integral kernel as 
\begin{equation*}
D^\alpha_{(p)} f(\xi) = \frac{1-p^\alpha}{1-p^{-\alpha-1}}\, \displaystyle\int_{\Qp} d\xi'\, 
\frac{f(\xi')-f(\xi)}{|\xi'-\xi|_p^{\alpha+1}}, \quad \alpha\in{\mathbb{C}} \label{VlaD}
\end{equation*}
They satisfy $D^{\alpha_1}_{(p)} D^{\alpha_2}_{(p)} = D^{\alpha_2}_{(p)} D^{\alpha_1}_{(p)} = D^{\alpha_1+\alpha_2}_{(p)}$.  

Since the roles of translation and phase are not going to be important in what follows, let us set $m=0$ and $l=1$ and define 
vectors $|n^{(p)}\rangle$ corresponding to $\psi^{(p)}_{-n+1,0,1}(\xi)$
\begin{equation}
\psi^{(p)}_{-n+1,0,1}(\xi)\: \longleftrightarrow\: |n_{(p)}\rangle  \label{WvletToVector}
\end{equation}
in the Hilbert space $L^2(\Qp)$. Then
\begin{align}
D_{(p)}^{\alpha} |n_{(p)}\rangle &= p^{n^{(p)}\alpha} |n_{(p)}\rangle\nonumber\\
\log_p D_{(p)} |n_{(p)}\rangle &= \displaystyle{\lim_{\alpha\to 0}}\, \frac{D_{(p)}^\alpha - 1}{\alpha\ln p}
|n_{(p)}\rangle\: = \: n_{(p)} |n_{(p)}\rangle \label{EigensystemVD}
\end{align}
The wavelets, by construction, transform naturally under the affine group of scaling and translation. However, it was shown in 
Ref.\cite{Dutta:2018qrv} that the scaling part of it enhances to a larger SL(2,$\mathbb{R})$ symmetry. In terms of the raising and 
lowering operators $a^{(p)}_\pm|n_{(p)}\rangle = |n_{(p)}\! \pm\! 1\rangle$ the generators of SL(2,$\mathbb{R}$) are 
$J^{(p)}_\pm = a_\pm^{(p)}\log_p\!D_{(p)}$ and $J_3^{(p)} = \log_p D_{(p)}$. The algebra of these generators and their action 
on the wavelet states are as follows.
\begin{align}
\begin{split}
\left[ J_3^{(p)},  J^{(p)}_\pm \right] = \pm J^{(p)}_\pm, &\qquad 
\left[ J_+^{(p)},  J^{(p)}_- \right] = -2 J^{(p)}_3\\ 
J^{(p)}_3 |n_{(p)}\rangle = n_{(p)} |n_{(p)} \rangle, &\qquad  
J^{(p)}_\pm |n_{(p)}\rangle = n_{(p)} |n_{(p)}\! \pm\! 1\rangle  
\end{split}\label{sl2}
\end{align}

We can now write \cref{EulerExpand} as
\begin{eqnarray}
\zeta(s) &=& \prod_{p\,\in\,{\mathrm{primes}}} \sum_{n^{(p)}=0}^\infty \left\langle n_{(p)} \right| 
D_{(p)}^{-s} \left| n_{(p)}\right\rangle \nonumber\\   
&=& \sum_{{\mathbf{n}}=(n_{(2)},n_{(3)},\cdots)}\!\!\!\! \left\langle{\mathbf{n}} \right| e^{-s \ln \mathcal{D}} 
\left| {\mathbf{n}}\right\rangle \label{ZetaAsTrace}
\end{eqnarray}
where we have used a shorthand $\ln \mathcal{D} \equiv \sum_p \ln p\, \log_p\! D_{(p)}$, and the vectors $\left| {\mathbf{n}}\right\rangle$ 
belong to the product of the Hilbert spaces for all primes $\bigotimes_p L^2(\Qp)$. However, since the sum runs only over the \emph{positive} 
integers (including zero), this subspace is actually $\bigotimes_p L^2(p^{-1}\Zp)$, spanned by the Bruhat-Schwarz functions restricted to 
$p^{-1}\Zp$ due to which the trace is well defined (see \cite{Kozyrev:2001,Dutta:2018qrv,WilsonBook} for details on the wavelet functions). 
This expression leads us to think of the zeta function as the partition function of a statistical system, the configurations of which are 
parametrised by the integers ${\mathbf{n}}=(n_{(2)},n_{(3)},\cdots)$.

The $\mathfrak{sl}_2(\mathbb{R})$ algebra \cref{sl2} can be realised in terms of a pair of oscillators in the Schwinger representation
\begin{equation}
J^{(p)}_3 = \log_p D_{(p)}  = \frac{1}{2}\left(N_{\mathrm{I}(p)} - N_{\mathrm{II(p)}} \right),\:\,
J^{(p)}_+ = a_{\mathrm{I}(p)}^\dagger a_{\mathrm{II}(p)}\:\, \text{ and }\:\, 
J^{(p)}_- = a_{\mathrm{II}(p)}^\dagger a_{\mathrm{I}(p)} 
\end{equation}
Formally there is a \emph{phase difference operator} $\Phi_{(p)} = \left(\Phi^{(p)}_{\mathrm{I}} - \Phi^{(p)}_{\mathrm{II}}\right)$ conjugate 
to the number difference operator $N_{(p)} = \frac{1}{2} \left(N_{\mathrm{I}(p)} - N_{\mathrm{II}(p)} \right)$, such that 
\begin{equation}
[\Phi_{I(p)}\, , N_{J(p')}] =  i \delta_{IJ}\delta_{pp'}, \qquad  [\Phi_{(p)}, N_{(p')}] =  i\delta_{pp'} \label{CCPhiN}
\end{equation}
In \cref{sec:Ising} we reviewed a construction for the phase operator following 
\cite{SusskindGlo,PhaseOpRevCN,GarrisonWong,GalindoPhase,PeggBarnett,BuschEtAl,MaRhodes}. 
Assuming for the moment that a phase operator with the desired properties can be constructed, we define the operator 
$\frac{1}{\ln p}\,\Phi_{(p)} p^{-N_{(p)}} = \frac{1}{\ln p}\,\Phi_{(p)} e^{-N\ln p}$ and evaluate the following commutator
\begin{equation}
\left[\frac{1}{\ln p}\,\Phi_{p}\, p^{-N_{(p)}}, {p'}^{N_{(p')}}\right] =  i \delta_{pp'} 
\label{Commutator}
\end{equation}
using \cref{CCPhiN}. Thus, the operator $\frac{1}{\ln p}\,\Phi_{(p)}\, {p}^{-N_{(p)}}$ is formally canonically conjugate to 
$p^{N_{(p)}} = D_{(p)}$.
 
We would now like to extend it to the large Hilbert space obtained by combining all primes. Let us first consider all prime numbers 
up to a fixed prime $\mathfrak{p}$. The number of such primes is $\pi(\mathfrak{p})$, where $\pi(x)$ is the prime counting function. 
We now define 
\begin{equation*}
\mathcal{O}_{\mathfrak{p}} = \frac{1}{\pi(\mathfrak{p})} \sum_{p=2}^{\mathfrak{p}} \frac{1}{\ln p}\,\Phi_{(p)}\,\mathcal{D}_{\mathfrak{p}}^{-1} 
\quad \text{ and } \quad 
\ln\mathcal{D}_{\mathfrak{p}} = {\sum_{p=2}^{\mathfrak{p}} \ln D_ {(p)}}   
\end{equation*}
which are operators in the truncated Hilbert space $\displaystyle{\bigotimes_{p=2}^{\mathfrak{p}}} L^2(p^{-1}\Zp)$. These are canonically 
conjugate since $\left[\mathcal{O}_{\mathfrak{p}},\mathcal{D}_{\mathfrak{p}}\right] = i$. Now we take the limit $\mathfrak{p}\to\infty$ 
to obtain the canonically conjugate operators
\begin{equation*}
\mathcal{O} = \lim_{\mathfrak{p}\to\infty} \mathcal{O}_{\mathfrak{p}},  \quad 
\mathcal{D} = \lim_{\mathfrak{p}\to\infty} \mathcal{D}_{\mathfrak{p}} \quad\text{such that}\quad \left[\mathcal{O},\mathcal{D}\right] = i  
\end{equation*}
on the large Hilbert space $\bigotimes_p L^2(p^{-1}\Zp)$. This limit is analogous to the \emph{thermodynamic limit} of statistical models, 
as we shall see in \cref{ssec:RZPhOp}.
 
Associated to these operators is the Weyl symmetric product 
\begin{align}
\begin{split}
\frac{1}{2} \left(\mathcal{D} \mathcal{O} \right. & + \left.\mathcal{O}  \mathcal{D}\right) 
= \lim_{\mathfrak{p}\to\infty} \frac{1}{2} \left(\mathcal{D}_{\mathfrak{p}} \mathcal{O}_{\mathfrak{p}}  + \mathcal{O}_{\mathfrak{p}}
\mathcal{D}_{\mathfrak{p}}\right) 
\:=\: \mathcal{O}  \mathcal{D} - \frac{i}{2}\\
&= \lim_{\mathfrak{p}\to\infty} \frac{1}{\pi(\mathfrak{p})}  \sum_{p=2}^{\mathfrak{p}} \left( 
\mathbf{1}\otimes\cdots\otimes e^{\frac{1}{2}\!\ln D_{(p)}} \frac{\Phi_{(p)}}{\ln p} e^{-\frac{1}{2}\!\ln D_{(p)}} 
\otimes\mathbf{1}\otimes\cdots\right)  
\end{split} \label{WeylOrd}
\end{align} 
which is (formally) self-adjoint. In the last line, we have a similarity transform of the sum of the $\Phi_{(p)}$ operators. As has been 
emphasised, e.g.\ in Ref.~\cite{GalindoPhase}, the operator canonically conjugate to the number operator can only be defined up to a 
similarity transformation. Hence there ought to be more than one (which could be infinite in number) \emph{total} phase operators $\Phi$ 
canonically conjugate to the \emph{total} number operator $\ln\mathcal{D} = \sum_{p} \ln D_{(p)}$. One may follow proposals in the 
literature (e.g. \cite{GalindoPhase}) to define $\Phi_{(p)}$, which would result in $\frac{1}{2} \left(\mathcal{D}_{\mathfrak{p}} 
\mathcal{O}_{\mathfrak{p}} + \mathcal{O}_{\mathfrak{p}}  \mathcal{D}_{\mathfrak{p}}\right)$, canonically conjugate to 
$\ln\mathcal{D}_{\mathfrak{p}}$, on the outer product of a dense subspace of the Hilbert space $L^2(p^{-1}\Zp)$ at the $p$-th place. It is 
worth reiterating that the construction discussed above is formal. The limit $\mathfrak{p}\to\infty$ is not straightforward. There is a more 
convenient way to construct the phase operator over a subspace of the Hilbert space. We shall attempt to do so in the next two subsections.

\subsection{Aggregate phase operator for the Riemann zeta function}\label{ssec:RZPhOp}
Let us return to the model of $\mathfrak{su}(2)$ spin in an external field of \cref{sec:Ising} with the Hamiltonian containing a site dependent 
magnetic field
\begin{equation*}
H = - \displaystyle{\sum_{p=2}^\mathfrak{p}}B_{p}N_{p} = - \displaystyle{\sum_{p=2}^\mathfrak{p}}B_{p}\left(S_{3,p} + j\mathbf{1}\right)
\end{equation*}
where we have now chosen an unusual convention\footnote{It should, however, be mentioned that this type of numbering has been used 
before in \cite{Spector1990,Julia1990,Bakas1991,Julia1994}.} of using \emph{prime numbers} $p$ to label the sites, $B_p$ are the values 
of the magnetic field at site $p$ and we have shifted the zero of the energy for convenience. The latter amounts to a shift in the spectrum of 
$S_{p,3}$ by $S_{p,3} \rightarrow N_p = S_{3,p} + j\mathbf{1}$ so that $N_{p}$ takes the integer value $0,1,\cdots, \mathfrak{n}$. 

In this case one can define a phase operator at an individual site, say $\boldsymbol{\phi}_p$ at the $p$-th site, as in \cref{sec:Ising}. Each 
of these individual operators satisfies the shift covariance relation (or commutator) for special values of $\beta$
\begin{align}
\big[ \boldsymbol{\phi}_p, e^{\beta\sum_{2}^{\mathfrak{p}}B_{p}N_{p}} \big] &= i \beta e^{\beta\sum_{2}^{\mathfrak{p}}B_{p}N_{p}}
\label{zetacovariance} \\
\text{for } \beta = \frac{2\pi ik}{B_{p}(\mathfrak{n}+1)} \pmod{{2\pi}/{B_{p}}}\: &\text{ with } k=1,\cdots,\mathfrak{n} \text{ and } 
p=2,\cdots,\mathfrak{p}
\nonumber
\end{align}
This is valid over the entire Hilbert space, i.e., on an arbitrary state vector, but only for these special values of $\beta$. Thus there are as 
many shift covariant phase operators as the number of sites, and each individual phase operator is covariant under the specific choices of 
$\beta$. Moreover, 
since each of the Hilbert spaces, labelled by $p$, is finite dimensional, the trace is a product over traces in each Hilbert space. Hence if 
we take the trace of \cref{zetacovariance}, exactly as in the case of the spin model in \cref{sec:Ising}, the trace of the commutator is zero, 
therefore, $\text{Tr}\, e^{\beta\sum_p B_pN_p} = 0$. Thus the shift covariance relation is valid for those values of $\beta$ which also satisfy 
the zero trace condition. This relates the zeroes of the partition function to the poles of the following resolvent operators 
\begin{equation*}
\mathfrak{R}[e^{i\boldsymbol{\phi}_p}](\phi) = \left(1-e^{-i\phi}e^{i\boldsymbol{\phi}_p}\right)^{-1}
\end{equation*}
for all $p$. The trace of the resolvent is
\begin{equation*}
\sum_{k_p=0}^{\mathfrak{n}}\frac{1}{1 - e^{-i\phi + i\phi_{k_p}}}
\end{equation*}
apart from the pole for $k=0$, which yields the trivial commutator.   

The similarity between the spin in a magnetic field and the zeta function is apparent at this stage. (Recall that we have labelled the edges 
connecting adjacent sites by the first $\mathfrak{p}$ prime numbers with this objective.) Indeed, if we choose the local magnetic field 
$B_{p} = \ln p$, then the partition function becomes
\begin{equation*}
Z(\beta) = \prod_{p=2}^{\mathfrak{p}}\bigg(\sum_{m_p=0}^{\mathfrak{n}} e^{\beta m_p \ln p}\bigg) = 
\prod_{p=2}^{\mathfrak{p}}\frac{1-p^{\beta(\mathfrak{n}+1)}}{1-p^{\beta}}
\end{equation*}
In the thermodynamic limit $\mathfrak{p}\to\infty$, even for {\em finite} $\mathfrak{n}$, the partition function has a simple form in terms of a 
ratio of the Riemann zeta functions
\begin{equation}
Z(\beta) = \lim_{\mathfrak{p}\to\infty}\, \prod_{p=2}^{\mathfrak{p}} \frac{1-p^{\beta(\mathfrak{n}+1)}}{1-p^{\beta}} = 
\frac{\zeta(-\beta)}{\zeta\left(-(\mathfrak{n}+1)\beta\right)} \label{PFasRatioRZeta}
\end{equation}
Remarkably this has the exact same form as the partition functions of a $\kappa$-parafermionic primon gas of 
\cite{Julia1990,Bakas1991,Julia1994,Schumayer:2011yp} with $\kappa=\mathfrak{n}+1$ and $s=-\beta$. It would be interesting to try to
relate the parafermionic variables to the spin degrees of freedom. Notice that $Z(\beta)$ has zeroes at the non-trivial zeroes of $\zeta(-\beta)$ 
from the numerator, as well as at $\beta = - 1/(\mathfrak{n}+1)$ from the pole of $\zeta\left(-(\mathfrak{n}+1)\beta\right)$ from the denominator. 
The latter is the only real zero, although it is at an unphysical value of the (inverse) temperature. However, the trivial zeroes of $\zeta(-\beta)$ 
are not zeroes of the partition function. This is due to the fact that at these points, both the numerator and the denominator have simple zeroes, 
hence 
\begin{equation*}
\lim_{\beta\to 2n} \frac{\zeta(-\beta)}{\zeta\left(-(\mathfrak{n}+1)\beta\right)} = \text{ finite}  
\end{equation*}
Thus the nontrivial zeroes are the Fisher zeroes of the spin model in the complex (inverse temperature) $\beta$-plane. However, since the 
zeroes of the Riemann zeta function are believed to be isolated (and since there is no accumulation point on the real line) these zeroes are 
not related to any phase transition. This is consistent as the system of spins in a magnetic field is not expected to undergo a phase transition.
Finally, the partition function has additional poles from the zeroes of the zeta function in the denominator.

The {\em spectrum of the zeroes} of the partition function is then given by
\begin{equation}
\sum_{p=2}^{\mathfrak{p}} \sum_{{n_p\in\mathbb{Z}}} \sum_{k_p=0}^{\mathfrak{n}} \frac{1}{1-e^{i\phi_{k_{p}} + 
i\frac{2\pi n_p}{\ln p}-i\phi}} - \sum_{p,n_p}\frac{1}{1-e^{i\frac{2\pi n_p}{\ln p}-i\phi}} \label{SpecZeroPF}
\end{equation}
where we have subtracted the pole due to $k=0$. This function may be rewritten as follows.
\begin{align*}
&\sum_{p=2}^{\mathfrak{p}}\sum_{n\in\mathbf{Z}} \frac{1}{1-e^{i\left(\frac{2\pi n}{(\mathfrak{n}+1)\ln p}-\phi\right)}} -
\sum_{p=2}^{\mathfrak{p}}\sum_{n\in\mathbf{Z}} \frac{1}{1-e^{i\left(\frac{2\pi n}{\ln p}-\phi\right)}}\\
&\buildrel{\text{sing.}}\over\approx \sum_{p=2}^{\mathfrak{p}} \sum_{n\in\mathbf{Z}} \frac{-i}{\phi - \frac{2\pi n}{(\mathfrak{n}+1)\ln p}}
- \sum_{p=2}^{\mathfrak{p}} \sum_{n\in\mathbf{Z}} \frac{-i}{\phi - \frac{2\pi n}{\ln p}}\\
&\approx \sum_{p=2}^{\mathfrak{p}}\frac{d}{d(i\phi)}\ln (1-p^{-(\mathfrak{n}+1)i\phi})-\frac{d}{d(i\phi)}\sum_{p=2}^{\mathfrak{p}}\ln (1-p^{-i\phi})\\
&\approx -i \frac{d}{d\phi} \ln \bigg(\prod_{p=2}^{\mathfrak{p}} \frac{ 1-p^{-(\mathfrak{n}+1)i\phi}}{1-p^{-i\phi}}\bigg)
\end{align*}
In the above we have used the Mittag-Leffler expansion, assuming analyticity of the partition function. The expression above, in the 
limit $\mathfrak{p}\to\infty$, becomes
\begin{equation*}
-i \frac{d}{d\phi}\ln\bigg(\frac{\zeta(i\phi)}{\zeta\left((\mathfrak{n}+1)i\phi\right)}\bigg)
\end{equation*}
for $\mathrm{Re}\,(i\phi) > 1$. 

\bigskip

We will now try to construct a {\em single} operator that can be understood as `canonically conjugate' to the Hamiltonian. If we define the 
total phase operator as $\boldsymbol{\Phi} = \sum_p \boldsymbol{\phi}_p$ (which is the sum of individual phase operators
$\boldsymbol{\phi}_p$ as defined in \cref{sec:Ising}) it does not, unfortunately, have the desired shift covariance relation 
\cref{IsingPhCommutator} with the Hamiltonian. This is due to the site dependence of the magnetic field $B_p$, as is apparent from the 
steps leading to \cref{IsingPhCommutator}. The commutator there is obtained only at specific discrete values of $\beta$ which are integer 
multiples of $2\pi k'/B_p(\mathfrak{n}+1)$. Therefore, unless the magnetic field $B_p$ at all the sites are commensurate, which is certainly 
not the case for $B_p = \ln p$, it is not possible to get the desired commutator this way. 

Instead, we propose to work with an {\em aggregate phase operator}  $\boldsymbol{\varphi}$ such that its action on the composite state 
$\bigotimes_p\big|\phi_{p,k_p}\big\rangle$ is defined to be $e^{i\hat{\phi}_p}$ if the eigenvalue $\phi_p \ne 0$, while all other eigenvalues
$\phi_{q\ne p}$ are zero, otherwise it acts as the identity. Thus, \emph{if two or more of the phases are non-zero}, $e^{i\boldsymbol{\varphi}} = 
\mathbf{1}$. This may be expressed as
\begin{equation}
e^{i\boldsymbol{\varphi}} =
\sum_{p=2}^{\mathfrak{p}} e^{i\hat{\phi}_p} \prod_{q\ne p} \delta_{\phi_q,0}  + \frac{2}{n_{\ne 0}(n_{\ne 0} - 1)}
\sum_{p_1,p_2=1}^{\mathfrak{p}} \prod_{p_1\ne p_2} (1-\delta_{\phi_{p_1},0})(1-\delta_{\phi_{p_2},0}) 
\label{AggregatePhOp}
\end{equation}
where $n_{\ne 0} = \sum_p (1 - \delta_{\phi_p,0})$ is the number of sites where the phase is non-zero. This is equivalent to projecting on a
subspace $\mathcal{H}^{(1)}$ of the Hilbert space, in which only one, and exactly one, phase is different from zero\footnote{This is analogous, 
though not exactly equivalent, to a projection of the Fock space of a quantum field theory of, say a scalar field, on a subspace with single-particle 
excitation.}. After the projection, one can use the total phase operator in the subspace $\boldsymbol{\Phi}|_{\mathcal{H}^{(1)}} = 
\Pi_{\mathcal{H}^{(1)}}\left(\sum_p \Phi_p\right) \Pi_{\mathcal{H}^{(1)}}$. {}From either point of view, the action of the above is nontrivial on a 
subspace of the Hilbert space parametrised by only one of the eigenvalues $\phi_p$ at a time, i.e., on a union of circles $\cup_p S^1_{(p)}$ 
while the full Hilbert space is parametrised by $(S^1)^{\mathfrak{p}}$. In the complement of this subspace, it is identity. In this subspace 
$\mathcal{H}^{(1)}$, we can follow the steps leading to \cref{IsingPhCommutator} to compute the commutator
\begin{equation*}
\left[ \boldsymbol{\varphi}, \Pi_{\mathcal{H}^{(1)}} e^{-\beta H} \Pi_{\mathcal{H}^{(1)}}\right] = i\beta\,\Pi_{\mathcal{H}^{(1)}}
e^{-\beta H} \Pi_{\mathcal{H}^{(1)}}
\end{equation*}
which holds in $\mathcal{H}^{(1)}$ for all $\beta = \frac{2\pi k}{B_p(\mathfrak{n}+1)}$ (mod $2\pi/ B_{p}$) where $k = 1,\cdots,\mathfrak{n}$ 
and $p=2,\cdots,\mathfrak{p}$. It is worth emphasising that, as in several examples in quantum theory, the domain of the canonical commutator 
is not the entire Hilbert space, but a direct sum of closed orthogonal subspaces\cite{GarrisonWong} of the type $\mathcal{H}^{(1)}$. In the limit 
$\mathfrak{p}\to\infty$, one take the closure of this subspace to obtain a closed subspace of the entire Hilbert space.

The resolvent function of the exponential of the aggregate phase operator \cref{AggregatePhOp} 
\begin{equation}
\mathfrak{R}[e^{i\boldsymbol{\varphi}}](\phi) =  \left(1 - e^{-i\phi} e^{i\boldsymbol{\varphi}}\right)^{-1}
\label{ResRZeta}
\end{equation}
has the trace
\begin{align}
\mathrm{Tr }\left(\mathfrak{R}[e^{i\boldsymbol{\varphi}}](\phi)\right)
& = \sum_{p=2}^{\mathfrak{p}} \sum_{k_1, \cdots, k_{\mathfrak{p}}} \left(\otimes_{i=1}^{\mathfrak{p}} 
\big\langle\phi_{k_i}\big|\right) e^{-\beta H} \sum_{n=0}^\infty e^{in(\boldsymbol{\varphi} - \phi)} \left(\otimes_{i=1}^{\mathfrak{p}} 
\big|\phi_{k_i}\big\rangle\right) \nonumber\\
&= \sum_{p=2}^{\mathfrak{p}} \Bigg(\sum_{{k_p\atop{\text{exactly one}\atop
\phi_{k_p} \ne 0}}} \sum_{n=0} e^{in\phi_{k_{p}}} e^{-in\phi} + \sum_{{k_p \atop{\text{at least two}\atop \phi_{k_p}\ne 0}}} 
e^{-in\phi}\Bigg) \nonumber\\
&= \, \sum_{p=2}^{\mathfrak{p}} \Bigg(\sum_{{k_p\atop{\text{exactly one}\atop\phi_{k_p} \ne 0}}}
\frac{1}{1-e^{i\phi_{k_{p}} - i\phi}} + \sum_{{k_p \atop{\text{at least two}\atop \phi_{k_p}\ne 0}}} \frac{1}{1-e^{-i\phi}}\Bigg)  
\label{RZResPoles}
\end{align}
Except for the pole at $\phi=0$, this behaviour is in fact identical to that of the resolvent $\displaystyle\sum_{p=2}^{\mathfrak{p}} 
\left(1 - e^{-i\phi} e^{i\hat{\phi}_{p}}\right)^{-1}$ in \cref{SpecZeroPF}.

Even though we do not require to take the limit $\mathfrak{n}\to\infty$, it is interesting to note that the phase operator $\boldsymbol{\phi}_p$ 
at the $p$-th site approaches the phase operator described in \cite{GalindoPhase}, which is a Toeplitz operator 
\cite{Gray2006,Widom1965,Nikolski2020} in this limit. This has been shown in \cite{BuschEtAl}. This provides a way to understand the relation 
to the spectrum without a truncation to a finite $\mathfrak{n}$ case. As shown in \cite{GalindoPhase,BuschEtAl}, each pair of operators 
$(\boldsymbol{\phi}_{p}, N_{p})$ satisfies the canonical commutation relation in a subspace $\Omega_{p}$ of the $p$-th Hilbert space 
$L^2(p^{-1}\Zp)$ as follows
\begin{equation*}
\Omega_p = \Big\{|f\rangle_{(p)} = \displaystyle\sum_{n_p=0}^{\infty} f_{n_p} |n_p\rangle_{(p)} : \displaystyle\sum_{n_p=0}^{\infty} 
 f_{n_p}=0\Big\}
\end{equation*} 
where, $|n_p\rangle_{(p)}$ is an eigenstate of $N_{p}$ corresponding to the eigenvalue $n_p$. Thus, each of the phase operators 
$\boldsymbol{\phi}_{p}$ satisfies the canonical commutator over a dense subspace of the Hilbert space
\begin{equation*}
\Big[\frac{1}{\ln p}\,\boldsymbol{\phi}_{p}\, , \sum_{p\in\text{prime}}\!\ln\mathcal{D}_{(p)}\Big] = i,\quad\text{in }\, 
\Omega_p\,\bigotimes_{p'\ne p} \mathcal{H}_{p'}
\end{equation*}
This is similar to the operator defined earlier, but without the sum restricted to a finite prime $\mathfrak{p}$ (along with the normalisation 
factor $\pi(\mathfrak{p})$ in the denominator). This is due to the fact that in this case, one gets a contribution from only one of the subspaces 
(one prime) at a time. In the next subsection we shall take a similar route to define another phase operator.

\subsection{Total phase operator for the Riemann zeta function}\label{ssec:AnotherPhOp}
Following \cite{GalindoPhase} (see also \cite{BuschEtAl}) we would like to discuss another construction of the phase operator $\hat{\Phi}$ 
conjugate to the number operator $\hat{N}$ such that
\begin{align}
\hat{N} |n\rangle &= n |n\rangle,\qquad n=0,1,\cdots,\mathfrak{n} \nonumber\\  
\hat{\Phi} &=  \sum_{m\ne n} \frac{i\, |m\rangle \langle n|}{m-n} \label{ToyNPhi}
\end{align}
This is a $(\mathfrak{n}+1)\times(\mathfrak{n}+1)$ hermitian Toeplitz matrix \cite{Gray2006,Widom1965,Nikolski2020}. When applied on a 
state $|\mathbf{v}\rangle = \sum_n v_n |n\rangle$, we find that 
\begin{equation}
\big[\hat{\Phi}, \hat{N}\big] |\mathbf{v}\rangle = i |\mathbf{v}\rangle\quad\text{if and only if } \sum_{n=0}^{\mathfrak{n}} v_n = 0 \label{ToyComm}
\end{equation}
Thus the commutator is valid in a codimension one subspace. For example, we could choose the $v_n$s to be the nontrivial 
$(\mathfrak{n}+1)$-th roots of unity. Toeplitz matrices and operators have a long history and have been studied extensively (see e.g., 
\cite{Nikolski2020}). Although eigenvalues $k_0 \le k_1 \le \cdots \le k_{\mathfrak{n}}$ and the corresponding eigenvectors $|k_m\rangle$ 
($k=0,1,\cdots,\mathfrak{n}$) of the matrix above exist, one cannot write them explicitly. Moreover, by Szeg\"{o}'s theorems, the spectrum 
is bounded by $\pi$ as $\mathfrak{n}\to\infty$ (so that the matrix size goes to infinity) and the eigenvalues are distributed uniformly and 
symmetrically around zero, as one can also check numerically for small values of $\mathfrak{n}$. 

\bigskip

Coming back to the problem of our interest, in which the Hamiltonian is $H = \sum_p \ln p\, N_{(p)}$, where $N_{(p)}$ is the number operator 
at the $p$-th site, which in turn can be expressed in terms of the generalised Vladimirov derivative. For a natural number $n\in\mathbb{N}$, 
we use the prime factorisation\footnote{Prime factorisation also plays an important role in the arithmetic gas models 
\cite{Spector1990,Julia1990,Bakas1991,Julia1994}.} to associate a vector in $\otimes_p L^2(p^{-1}\Zp)$ as 
\begin{equation}
n = \prod_p p^{n_{(p)}} \: \longleftrightarrow\: |\mathbf{n}\rangle = \otimes_p |n_{(p)}\rangle \label{factorKet}
\end{equation}
using the wavelet basis. We emphasise that only a finite number of entries in the infinite component vector are non-zero integers. Clearly 
$|\mathbf{n}\rangle$ is an eigenvector of $H$
\begin{equation*}
H |\mathbf{n}\rangle = \sum_p n_{(p)} \ln p\, |\mathbf{n}\rangle = \ln n\, |\mathbf{n}\rangle  
\end{equation*}
Moreover, these states are orthonormal $\langle \mathbf{n}_i | \mathbf{n}_j \rangle = \prod_p \langle n^i_{(p)} | n^j_{(p)} \rangle =
\prod_p \delta_{n^i_{(p)}n^i_{(p)}} = \delta_{n_i n_j}$. 

When restricted to a fixed value of $p$, the following definition for the phase operator 
\begin{equation*}
\frac{i}{\ln p} \sum_{n_a^{(p)} \ne n_b^{(p)}} \frac{|\mathbf{n}^{(p)}_a\rangle\, \langle\mathbf{n}^{(p)}_b|}{(n^{(p)}_a - n^{(p)}_b)} 
\end{equation*}
on $L^2(p^{-1}\Zp)$ is natural. This is a Toeplitz matrix, therefore, it has eigenvectors $|k_{(p)}\rangle$.  Let us define the phase operator on 
the full space $\otimes_p L^2(p^{-1}\Zp)$ schematically to be of the form
\begin{equation*}
\Phi_{\text{tot}} \sim  \sum_{n_a\ne n_b} \frac{i\, |\mathbf{n}_a\rangle \langle\mathbf{n}_b|}{\ln n_a - \ln n_b}
= \sum_{\stackrel{\text{not all}}{n_a^{(p)} = n_b^{(p)}}} \frac{i \left(\otimes_{p_a}|\mathbf{n}^{(p_a)}_i\rangle\right) \left(\otimes_{p_b}
\langle\mathbf{n}^{(p_b)}_b|\right)}{\sum_p (n^{(p)}_a - n^{(p)}_b)\ln p} 
\end{equation*}
We need to specify the limits of sums over the integers in the above. However, before we undertake that exercise, we would like to check if 
$H$ and $\Phi_{\text{tot}}$ could be a canonically conjugate pair, possibly on a subspace spanned by vectors of the form in \cref{factorKet}. 

To this end, let us now consider a {\em finite} linear combination of the form $|\mathbf{v}\rangle = \sum_{\mathbf{n}} v_{\mathbf{n}} 
|\mathbf{n}\rangle$, in which we further require the coefficients to factorise as $v_{\mathbf{n}} \equiv v_{(n_{(2)},n_{(3)},\cdots)} = \prod_p 
v_{n_{(p)}}$. One can compute the commutator and verify that on such a state
\begin{equation}
\big[\Phi_{\text{tot}}, H \big] |\mathbf{v}\rangle = i |\mathbf{v}\rangle\quad\text{if and only if } \sum_{n} v_n = 0
\label{PhiHComm}
\end{equation}
where the upper limit of the sum is the maximum integer $n_{\text{max}}$ that appear in the definition of the vector $|\mathbf{v}\rangle$.
Consider all the vectors $|n\rangle$ that appear in the linear combination in defining $|\mathbf{v}\rangle$ on which we want to check for
the commutator, and the prime factorisations of the corresponding integers $n$. Let the maximum $\text{max}_p \{n^{(p)}\}$ of these be
$\mathfrak{n}\in \mathbb{N}$. There is also a highest prime $\mathfrak{p}$, i.e., above which all $n^{(p>\mathfrak{p})}=0$ in the factorisations. 
We can now make the proposal for the phase operator more precise. It is 
\begin{equation}
\Phi_{\text{tot}} =\begin{cases}
\displaystyle{\sum^{\mathfrak{n}}_{{n_a^{(p)},\,n_b^{(p)}=0}\atop{\text{not all } n_a^{(p)} = n_b^{(p)}}} \!\!\!\!
\frac{i\left(\otimes_{p_a\le\mathfrak{p}} \big|\mathbf{n}^{(p_a)}_i\big\rangle\right) \left(\otimes_{p_b\le\mathfrak{p}}\big\langle\mathbf{n}^{(p_b)}_b
\big|\right)}{\sum_{p\le\mathfrak{p}} (n^{(p)}_a - n^{(p)}_b)\ln p}} & \mbox{for } n_a^{(p)},n_b^{(p)} \le \mathfrak{n}\\
\displaystyle{\bigotimes_{p>\mathfrak{p}} \big|n_{(p)}=0\big\rangle \big\langle n_{(p)}=0\big|} & \mbox{otherwise}
\end{cases}
\label{TotPhase}
\end{equation}
and acts on a space spanned by vectors of the form
\begin{equation*}
|\mathbf{k},\mathfrak{p}\rangle = \left(\otimes_{p \le \mathfrak{p}} |k_{(p)}\rangle\right) \otimes \left(\otimes_{p' > \mathfrak{p}}
|n_{(p')}=0\rangle \right)
\end{equation*}
where at least one $k_{(p)}\ne 0$ for a prime $p \le \mathfrak{p}$ and for $p>\mathfrak{p}$, we have chosen the `vacuum' state in the number
representation. In the limit $\mathfrak{p}\to\infty$ (even for finite values of $\mathfrak{n}$), we expect this to be a well defined Toeplitz operator 
on $\otimes_p L^2(p^{-1}\Zp)$. However, we are not able to offer a rigorous mathematical proof of this assertion.  

\bigskip

The phase operator cannot be defined uniquely, it is ambiguous upto a similarity transform \cite{GalindoPhase}. Given a phase operator $\Phi$, 
for example, as defined in \cref{ToyNPhi}, let us consider the operator $\Phi_\beta = e^{-\beta N}\Phi  e^{\beta N}$ related by a similarity transform 
labelled by a parameter $\beta$. This is would have been a trivial statement had the commutator \cref{ToyComm} been true in the full vector space, 
however, as we have seen this relation holds in a subspace of codimension one. It is straightforward to check that the condition that restricts to the 
subspace is modified to $\sum_n e^{\beta n} v_n = 0$ for $\Phi_\beta$ to be conjugate to $N$. We may choose 
$v_n = e^{2\pi i m_1 n/(\mathfrak{n}+1)}$ and $\beta = 2\pi i m_2/(\mathfrak{n}+1)$ with $m_1+m_2\ne 0\pmod {\mathfrak{n}+1}$. This condition 
is identical to the vanishing of the partition function $Z=\text{Tr } e^{-\beta H}$ for the Hamiltonian $H=-N$ at these special values of $\beta$.

Now consider $\Phi_{\text{tot},\beta} = e^{-\beta H}\Phi_{\text{tot}} e^{\beta H}$, the similarity transformation of \cref{TotPhase}. The modified 
condition that defines the subspace is $\prod_p \sum_{n_{(p)}} e^{\beta n_{(p)}\ln p} v_{n_{(p)}} = 0$. If we choose the coefficient 
$v_{n_{(p)}} = \chi(p)^{n_{(p)}}$, where $\chi(p)$ is a Dirichlet character (see \cref{DirichletL}), the subspace is defined by the vanishing of 
\begin{equation}
\prod_{p=2}^{\mathfrak{p}} \sum_{n_{(p)}=0}^{\mathfrak{n}} \left(\chi(p) p^\beta\right)^{n_{(p)}}  
= \prod_{p=2}^{\mathfrak{p}} \frac{1-\chi(p^{\mathfrak{n}+1})p^{\beta (\mathfrak{n}+1)}}{1-\chi(p)p^{\beta}} \label{SubspaceCondition}
\end{equation} 
which, in the limit $\mathfrak{p}\to\infty$ is a ratio of Riemann zeta or Dirichlet $L$-functions, depending whether the character
is trivial or not, as in \cref{PFasRatioRZeta,PFasDLFn}, respectively. Thus the subspace in which the phase operator \cref{TotPhase}, or its 
similarity transform, is canonically conjugate to the Hamiltonian is defined by the vanishing of the Riemann zeta function (at special values 
of the inverse temperature $\beta\ne0$). We have previously encountered this in \cref{PFasRatioRZeta} with the aggregate phase operator 
defined in \cref{ssec:RZPhOp}. As we see, different choices for the coefficients relate to the vanishing of Dirichlet $L$-functions, to which we 
shall now turn our attention.

\section{Extension to the Dirichlet $L$-functions}\label{sec:DirichletL}
The Riemann zeta function belongs to a family of functions, called the Dirichlet $L$-functions, that are defined as the analytic
continuation of the Dirichlet series 
\begin{equation}
L(s,\chi) = \sum_{n=1}^\infty \frac{\chi(n)}{n^s} = \prod_{p\,\in\,{\mathrm{primes}}} \frac{1}{1 - \chi(p) p^{-s}},
\qquad \mathrm{Re}(s) > 1 
\label{DirichletL}
\end{equation} 
to the complex $s$-plane. In the above, $\chi_(n)$, called the Dirichlet character, is a homomorphism from the multiplicative group
$G(k)=\left(\mathbb{Z}/k\mathbb{Z}\right)^*$ of \emph{invertible} elements of $\mathbb{Z}/k\mathbb{Z}$ to $\mathbb{C}^*$, which   
is then extended as a character for all $\mathbb{Z}$ by setting $\chi(m)=0$ for all $m$ which are zero $\pmod k$ \cite{Serre:Course}. 
A Dirichlet character so defined satisfy the following properties
\begin{enumerate}
\item
For all $m_1, m_2 \in \mathbb{Z}$, $\chi(m_1 m_2) = \chi(m_1) \chi(m_2)$
\item
$\chi(m) \ne 0$ if and only if $m$ is relatively prime to $k$
\item
$\chi(m_1) = \chi(m_2)$ if $m_1 \equiv m_2$ $\pmod k$
\end{enumerate}
Therefore, $\chi$ is a multiplicative character, defined modulo $k$, on the set of integers.  It is this multiplicative property that justifies
the sum to be written as an infinite product in \cref{DirichletL}. 

There is a trivial character that assigns the value 1 to all integers, including 0. (This may be taken to correspond to $k=1$.) The Riemann
zeta function corresponds to the choice of the trivial character. In all other cases, only those integers (respectively, primes), the Dirichlet 
characters of which are not zero, contribute to the sum (respectively, the product). This of course depends on the periodicity $k$ of the 
character. Therefore, the product restricts to primes that do not divide $k$
\begin{equation*}
\prod_{p} \frac{1}{1-\chi_{k}(p)p^{-s}} = \prod_{(p,k)=1} \frac{1}{1-\chi_{k}(p)p^{-s}}
= \prod_{(p,k)=1} \sum_{n_{p}=0}^{\infty}\chi_{k}(p)^{n_{p}}p^{-n_{p}s} 
\end{equation*}
With this understanding, one can define the \emph{inverse} $\chi^{-1}$ by restricting to the relevant set of primes. For these primes 
$p\nmid k$, the Dirichlet character satisfies $\chi_k^{-1}(p) = \chi_k^*(p)$. (Formally, for the others, we may take $\chi$ as well as 
$\chi^{-1}$ to be zero.)

Everything we discussed in the context of the Riemann zeta function in \cref{sec:RiemannZ}, including all the caveats, apply to the Dirichlet 
$L$-functions, with obvious modifications at appropriate places. The role of the generalised Vladimidrov derivative, acting on complex valued 
functions on the $p$-adic numbers $\Qp$ is played by the generalised Vladimirov derivative \emph{twisted by the character $\chi$} 
\cite{Dutta:2020qed}, be denoted by $D_{(p)\mathfrak{x}}$. The Kozyrev wavelets are eigenfunctions of these operators for all $\chi$. 
The eigenvalues, however, are different and  involve the Dirichlet character as follows
\begin{equation}
D_{(p)\mathfrak{x}} \psi_{1-n,m,j}^{(p)}(\xi) = \chi_{k}(p^{n})p^{n}\, \psi_{1-n,m,j}^{(p)}(\xi) \label{TwistedVlaD}
\end{equation}
We refer to \cite{Dutta:2020qed} for details of the construction and other properties of these operators.

The above equation and \cref{VDonKozy} lead to the conclusion that $D$ and $D_{\mathfrak{x}}$ are simultaneously diagonalisable, 
hence the Kozyrev wavelets are also eigenfunctions of the \emph{unitary} operator $U_{\mathfrak{x}} = D_{\mathfrak{x}} D^{-1}$
\begin{equation}
U_{(p)\mathfrak{x}}\psi_{1-n,m,j}^{(p)}(\xi) = D_{(p)\mathfrak{x}} D^{-1}_{(p)} \psi_{1-n,m,j}^{(p)}(\xi) = 
\chi_{k}(p^{n})\psi_{1-n,m,j}^{(p)}(\xi) \label{DXDinv}
\end{equation}
We can define its inverse $U_{(p)\mathfrak{x}}^{-1} = U_{(p)\mathfrak{x}}^\dagger = D_{(p)} D^{-1}_{(p)\mathfrak{x}^{*}}$ for those $k$ which 
do not contain $p$ in its factorisation (otherwise it is the identity operator). Conversely, when we consider all primes, for a given $k$, we need 
to restricted to the set of primes that do not divide $k$, i.e., with the formal extension of the inverse given after \cref{EulerL}.

As in the case of the Riemann zeta function, we can combine all the prime factors to write $L(s,\chi_{k})$ as a trace
\begin{equation*}
L(s,\chi) = \sum_{{\mathbf{n}} = (n^{(2)},n^{(3)},\cdots)}\!\!\!\! \big\langle{\mathbf{n}} \big| \,
\mathcal{U}_{\mathfrak{x}} e^{-s\ln\mathcal{D}} \,\big|\mathbf{n}\big\rangle
\end{equation*}
where $\mathcal{U}_{\mathfrak{x}}= \otimes_{p}U_{(p)\mathfrak{x}}$. If we want to interpret this as the partition function of a statistical 
mechanical model, the Hamiltonian is such that 
\begin{equation*}
e^{-\beta H_{\mathfrak{x}}}\: \longleftrightarrow\: \mathcal{U}_{\mathfrak{x}} e^{-s\ln\mathcal{D}} =  \mathcal{D}^{-1} 
\mathcal{D}_{\mathfrak{x}} \mathcal{D}^{-s} = \mathcal{D}_{\mathfrak{x}} \mathcal{D}^{-s-1}   
\end{equation*}
which reduces to $e^{-\beta H} \sim \mathcal{D}^{-s}$ corresponding to the Riemann zeta function in \cref{ZetaAsTrace}, upto a phase. Now 
since a non-zero $\chi_{k}(p) = e^{i\omega_{p}}$ is a root of unity, we can define a new phase state\footnote{This is done by truncating the 
spectrum to relate with the previous case.}
\begin{equation}
|\phi_{k_{p}}^{(\mathfrak{x})} \rangle = \frac{1}{\sqrt{\mathfrak{n}+1}} \sum_{n_p=0}^{\mathfrak{n}} 
e^{-i n_p (\phi_{k_{p}}\ln p + \omega_p)} | n_p \rangle  \label{LFnPhState}
\end{equation}
which provide an orthonormal set 
\begin{equation*}
\big\langle \phi_{k'_{p}}^{(\mathfrak{x})} \big| \phi_{k_{p}}^{(\mathfrak{x})} \big\rangle = \delta_{k_{p},k'_{p}}
\end{equation*}
One may construct a phase operator
\begin{equation}
\hat{\phi}_{(p)\mathfrak{x}} = \sum_{k_{p}} \left( \phi_{k_{p}} + \frac{\omega_{p}}{\ln p} \right) \big| \phi_{k_{p}}^{(\mathfrak{x})} \big\rangle
\big\langle \phi_{k_{p}}^{(\mathfrak{x})} \big| \;
\equiv\: \sum_{k_{p}} \phi_{k_{p}}^{(\mathfrak{x})} \big| \phi_{k_{p}}^{(\mathfrak{x})} \big\rangle
\big\langle \phi_{k_{p}}^{(\mathfrak{x})} \big| \label{LFnPhOp} 
\end{equation}
using the eigenvalues and eigenstates as before.

Now we define the operator $U_{(p)}e^{\beta \ln p N_p}$ such that
\begin{equation}
U_{(p)} e^{\beta \ln p N_p} |n_p \rangle = e^{i n_p\omega_p} e^{\beta n_p \ln p} |n_p \rangle
\label{twistedNOp}
\end{equation}
It follows that
\begin{align*}
\hat{\phi}_{(p)\mathfrak{x}} U_{(p)} e^{\beta \ln p\, N_p}  
&= U_{(p)} e^{\beta \ln p\, N_p} \, \sum_{k_{p}} \left(\phi_{k_p}^{(\mathfrak{x})} + \frac{\omega_p}{\ln p} - i \beta\right) 
\Big| \phi_{k_{p}}^{(\mathfrak{x})}+\frac{\omega_{p}}{\ln p} - i\beta \Big\rangle \Big\langle \phi_{k_{p}}^{(\mathfrak{x})} + 
\frac{\omega_{p}}{\ln p} - i\beta \Big| \nonumber\\
&{}\quad + U_{(p)}e^{\beta \ln p\, N_{p}} \left(i\beta-\frac{\omega_p}{\ln p}\right) \sum_{k_{p}} \Big| \phi_{k_{p}}^{(\mathfrak{x})}+
\frac{\omega_{p}}{\ln p} - i\beta \Big\rangle \Big\langle \phi_{k_{p}}^{(\mathfrak{x})}+\frac{\omega_{p}}{\ln p} - i\beta \Big| 
\end{align*}
This relation can be obtained by the same method used in the earlier sections. If $i\beta$ takes any of the values 
$\frac{2\pi k_p}{(\mathfrak{n}+1)\ln p} + \frac{\omega_p}{\ln p}$ then in the first term above, one gets the phase operator \cref{LFnPhOp}, 
since $\phi_{k_p}$ is defined modulo $\frac{2\pi}{\ln p}$. Hence, as in the case of the Riemann zeta function, 
\begin{equation}
\Big[ \hat{\phi}_{(p)\mathfrak{x}} , U_{(p)\mathfrak{x}} e^{\beta \ln p\, N_p}\Big] = \left(i\beta-\frac{\omega_p}{\ln p}\right) 
U_{(p)\mathfrak{x}} e^{\beta \ln p\, N_p} \label{TwistedPhiNCommutator}
\end{equation}
The definition of the (exponential of the) resolvent is completely analogous to the case of the Riemann zeta function \cref{ResRZeta} --- 
one only needs to substitute $\hat{\phi}_{p}\to\hat{\phi}_{(p)\mathfrak{x}}$, resulting in the trace
\begin{equation*}
\sum_{p=2}^{\mathfrak{p}} \Bigg( \sum_{{k_p\atop{\text{exactly one}\atop\phi_{k_p} \ne 0}}} \frac{1}{1 - e^{i\left(\phi_{k_p} + 
\frac{\omega_p}{\ln p}\right) - i\phi}}\; + \sum_{{k_p \atop{\text{at least two}\atop \phi_{k_p}\ne 0}}} 
\frac{1}{1-e^{\frac{i\omega_p}{\ln p}-i\phi}} \Bigg)
\end{equation*}
in place of \cref{RZResPoles}. Once again the poles (apart from that at $\phi=0$) coincide with the zeroes of the partition function, which 
is\footnote{The unitary opeartor $\mathcal{U}_{\mathfrak{x},\mathfrak{p}}$ is the product of the corresponding operators at all sites 
$\mathcal{U}_{\mathfrak{x},\mathfrak{p}} = \displaystyle{\prod_{p=1}^{\mathfrak{p}}} U_{(p)\mathfrak{x}}$.}
\begin{equation}
Z(\beta) = \mathrm{Tr } \left(\mathcal{U}_{\mathfrak{x},\mathfrak{p}} \exp\Big(\beta \sum_{p=2}^{\mathfrak{p}} \ln p\, N_p\Big)\right)
= \prod_{p=2}^{\mathfrak{p}} \frac{1-\chi^{\mathfrak{n}+1}(p)p^{\beta (\mathfrak{n}+1)}}{1-\chi(p)p^{\beta}} \label{LPFfinitep} 
\end{equation}
where we have used the fact that $\chi(p^{\mathfrak{n}+1})(p)=\chi^{\mathfrak{n}+1}(p)$ is again a character with the same periodicity. 
In the thermodynamic limit  $\mathfrak{p} \to \infty$ we get the following ratio of the Dirichlet $L$-functions
\begin{equation}
Z(\beta) = \frac{L(-\beta,\chi_k)}{L(-(\mathfrak{n}+1)\beta,\chi^{\mathfrak{n}+1}_{k})} \label{PFasDLFn}
\end{equation}
In the special case where $\mathfrak{n}+1$ is the Euler totient function $\varphi(k)$ or its integer multiple, $\chi(p^{\varphi(k)})=
\left(\chi(p)\right)^{\varphi(k)}=\chi_{k,0}(p)$, the principal character, which is 1 if $(p,k)=1$ and 0 otherwise.
Except for the trivial zeroes, the script of the discussions above is very similar to what we argued for the Riemann zeta function.

\bigskip

In summary, we have proposed to view the Riemann zeta and the Dirichlet $L$-functions as the partition functions (upto multiplication by a 
function that plays no essential role) of \emph{quantum spins} in magnetic fields, the values of which depend on the site. We have argued 
how to make sense of the phase operator (upto a similarity transformation). The zeroes of the partition function coincide with the poles of 
the resolvent function of the exponential of the aggregate or total phase operators, as discussed in \cref{ssec:RZPhOp,sec:DirichletL}. A
different approach to the phase operator was discussed in \cref{ssec:AnotherPhOp}. Its relation to the partition function, via similarity 
transforms, seems to relate the zeta function and the $L$-functions in the same framework. 


\bigskip

\noindent{\bf Acknowledgements:} We thank Surajit Sarkar for collaboration at initial stages of this work. It is a pleasure to acknowledge 
useful discussions with Rajendra Bhatia and Ved Prakash Gupta. We would like to thank Toni Bourama and Wilson Z\`{u}\~{n}iga-Galindo
for the invitation to write this article.

\bibliographystyle{hieeetr}
\bibliography{PseuD}{}

\end{document}